\documentclass[twocolumn,preprintnumbers,amsmath,amssymb,showpacs,prl,apsrev4-1]{revtex4-1}
\usepackage{graphicx}
\usepackage{dcolumn}
\usepackage{bm}
\usepackage{SIunits}
\usepackage{verbatim}
\usepackage{placeins}

\begin{document}
\title{Interplay of iron and rare-earth magnetic order in rare-earth iron pnictide superconductors under magnetic field}
\author{Lei-Lei Yang$^{1}$,
        Da-Yong Liu$^{1}$,
        Dong-Meng Chen$^{2}$,
    and Liang-Jian Zou$^{1,3}$}
\altaffiliation{Corresponding author} \email{dyliu@theory.issp.ac.cn; zou@theory.issp.ac.cn}
\affiliation{
      $^1$ Key Laboratory of Materials Physics, Institute of Solid State
      Physics, Chinese Academy of Sciences, P. O. Box 1129, Hefei, Anhui
      230031, China \\
      $^2$ College of Science, China University of Petroleum, Qing Dao, 266580, China \\
      $^3$ Department of Physics, University of Science and Technology of China, Hefei, 230026, China
}

\date{\today}

\begin{abstract}
The magnetic properties of iron pnictide superconductors with magnetic rare-earth ions under strong magnetic field are investigated based on the cluster self-consistent field method. Starting from an effective Heisenberg model, we present the evolution of magnetic structures on magnetic field in RFeAsO (R=Ce, Pr, Nd, Sm, Gd and Tb) and RFe$_{2}$As$_{2}$ (R=Eu) compounds. It is found that spin-flop transition occurs in both rare-earth and iron layers under magnetic field, in good agreement with the experimental results. The interplay between rare-earth and iron spins plays a key role in the magnetic-field-driven magnetic phase transition, which suggests that the rare-earth layers can modulate the magnetic behaviors of iron layers. In addition, the factors that affect the critical magnetic field for spin-flop transition are also discussed.
\end{abstract}


\vskip 300 pt

\maketitle

\section{Introduction}
Since the iron-based superconductors were discovered~\cite{JACS130-3296}, the families of iron pnictides and chalcogenides display very rich phase diagrams, including antiferromagnetism or spin-density wave~\cite{EPL83-27006,Nature453-899}, superconductivity~\cite{JACS130-3296}, structural/magnetic phase transitions~\cite{Nature453-899}, orbital ordering~\cite{PRL103-267001,PRB81-180514,PRL104-057002,PRB84-064435} and nematic ordered phase~\cite{PRB77-224509}, {\it etc.}. The proximity of magnetism and superconductivity suggests the spin degree of freedom plays a key role in understanding the basic low-energy physics of the iron-based superconductors. And spin fluctuations have been proposed as the unconventional superconducting mechanism of iron-based superconductors~\cite{PRL101-087004}. Thus, to understand the magnetic properties is a primary task in uncovering the microscopic mechanism of the iron-based superconductivity.

Through many magnetic measurements, mainly with the help of the neutron scattering techniques, various antiferromagnetic (AFM) structures in different ironpnictide compounds have been determined;
the 1111 (RFeAsO with R=rare-earth ions), the 111 (AFeAs with A=alkali metal ions, {\it e.g.} Na) and the 122 phases (AFe$_{2}$As$_{2}$ with A=alkaline earth metal ions, {\it e.g.} Ba, and RFe$_{2}$As$_{2}$ with R=rare-earth ions, {\it e.g.} Eu), possess striped AFM (SAFM) order~\cite{Nature453-899,PRL101-257003,PRL102-227004,PRB80-174424}, while FeTe is bi-collinear AFM (BAFM)~\cite{PRL102-247001}, {\it etc.}. Among these compounds, the parent compounds with magnetic rare-earth ions, such as RFeAsO (R=Ce, Pr, Nd, Sm, Gd and Tb) and EuFe$_{2}$As$_{2}$, have demonstrated particular interesting. SmFeAsO$_{1-x}$F$_{x}$ shows the maximum T$_{c}$ with about 55 K~\cite{Nature453-761}, in other iron-pnictides with magnetic rare-earth ions, the T$_{c}$ is about 41 K for Ce~\cite{nmat7-953}, 52 K for Pr~\cite{PRB78-132504}, and 51 K for Nd~\cite{EPL82-57002}. However, in EuFe$_{2}$As$_{2}$, the T$_{c}$ is only about 29.5 K~\cite{PRB79-212509}. Since the basic FeAs units are very similar in these systems, the superconducting transition temperatures T$_{c}$ are very different, indicating a possibly distinct magnetic interaction between rare-earth and Fe layers in different compounds, and these distinct magnetic coupling is crucial in promoting the superconducting transition temperature.

Especially, these magnetic rare-earth ions exhibit different magnetic interactions, resulting in a more complicated magnetic behaviors. For example, in 1111 system, the rare earth ions of RFeAsO are AFM, and undergo an AFM-paramagnetic phase transition at T$_{N}$~\cite{PRB78-052502}. While in 122 system, such as in EuFe$_{2}$As$_{2}$, Eu$^{2+}$ ions display a ferromagnetic (FM) ordering~\cite{PRB80-174424}, {\it etc.}. A series of experiments had been performed to investigate the role of the magnetic rare-earth ions on the Fe-3$d$ magnetism and superconductivity.
Meanwhile, the influence of magnetic field on these complicated magnetic ordering are also investigated using static and pulsed field techniques. It is found when a magnetic field is applied in EuFe$_{2}$As$_{2}$, a spin flop transition occurs in rear-earth Eu layer observed experimentally at a very low magnetic field~\cite{NJP11-025007,PRB81-220406,JLTP159-601}. In addition, in SmFeAsO, at a high pulsed magnetic field, a spin-flop like transition is also observed~\cite{PRB83-134503}.

These experiments have demonstrated that in ironpnictide superconductors the magnetic structures under magnetic field display a complex phenomenon. Thus, a question naturally arises: what role do the magnetic rare-earth ions and magnetic field play on the magnetism and superconductivity of the FeAs layers in these rare-earth compounds? To address this question, a detail theoretical investigation is expected.
In this paper, we present the effect of magnetic rare-earth ions and magnetic field on the magnetism of iron pnictide compounds and investigate the interplay of magnetic rear-earth and Fe ions. This paper is organized as follows: a model Hamiltonian and the cluster self-consistent field (Cluster-SCF) method are described in Sec.II; then the results and discussions are presented in Sec. III; the last section is devoted to the remarks and conclusions.

\section{Model Hamiltonian and Method}

It has been shown that the effective Heisenberg models provide a reasonable description for the magnetic structure and spin wave behaviors in the iron-based superconductors~\cite{PRB42-6861}. The $J_{1}$-$J_{2}$ frustrated Heisenberg model was firstly proposed to describe the magnetic properties of the iron-based superconductors. While, the inelastic neutron scattering experiments found a large in-plane anisotropy in the magnetic interactions~\cite{nphys5-555}, suggesting the magnetic exchange constants $J_{1a} > J_{1b}$ with $a$ denoting the AFM direction and $b$ the FM direction. Thus an effective $J_{1a}$-$J_{1b}$-$J_{2}$ model could well describe the striped AFM (SAFM) order in iron-based compounds~\cite{PRB81-024505}. Here we start from the $J_{1a}$-$J_{1b}$-$J_{2}$ Heisenberg model.
\begin{figure}[htbp]
\centering
\hspace*{-9mm}
\begin{minipage}[t]{0.45\linewidth}
    \centering
    \includegraphics[trim = 0mm 0mm 0mm 0mm, clip=true, width=1.45in]{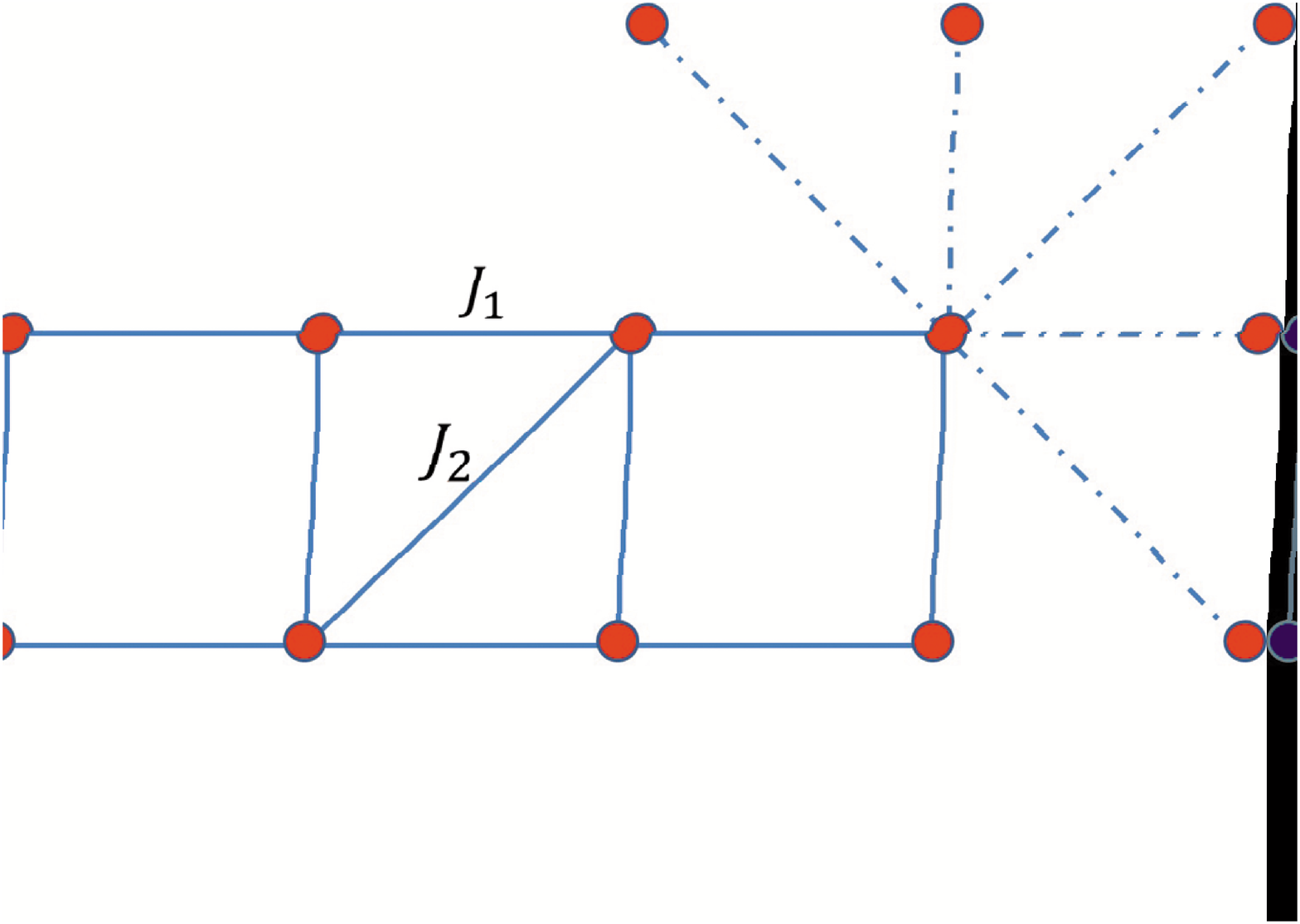}
\end{minipage}
\hspace{0.4ex}
\begin{minipage}[t]{0.45\linewidth}
    \centering
    \includegraphics[trim = 0mm 0mm 0mm 0mm, clip=true, width=1.45in]{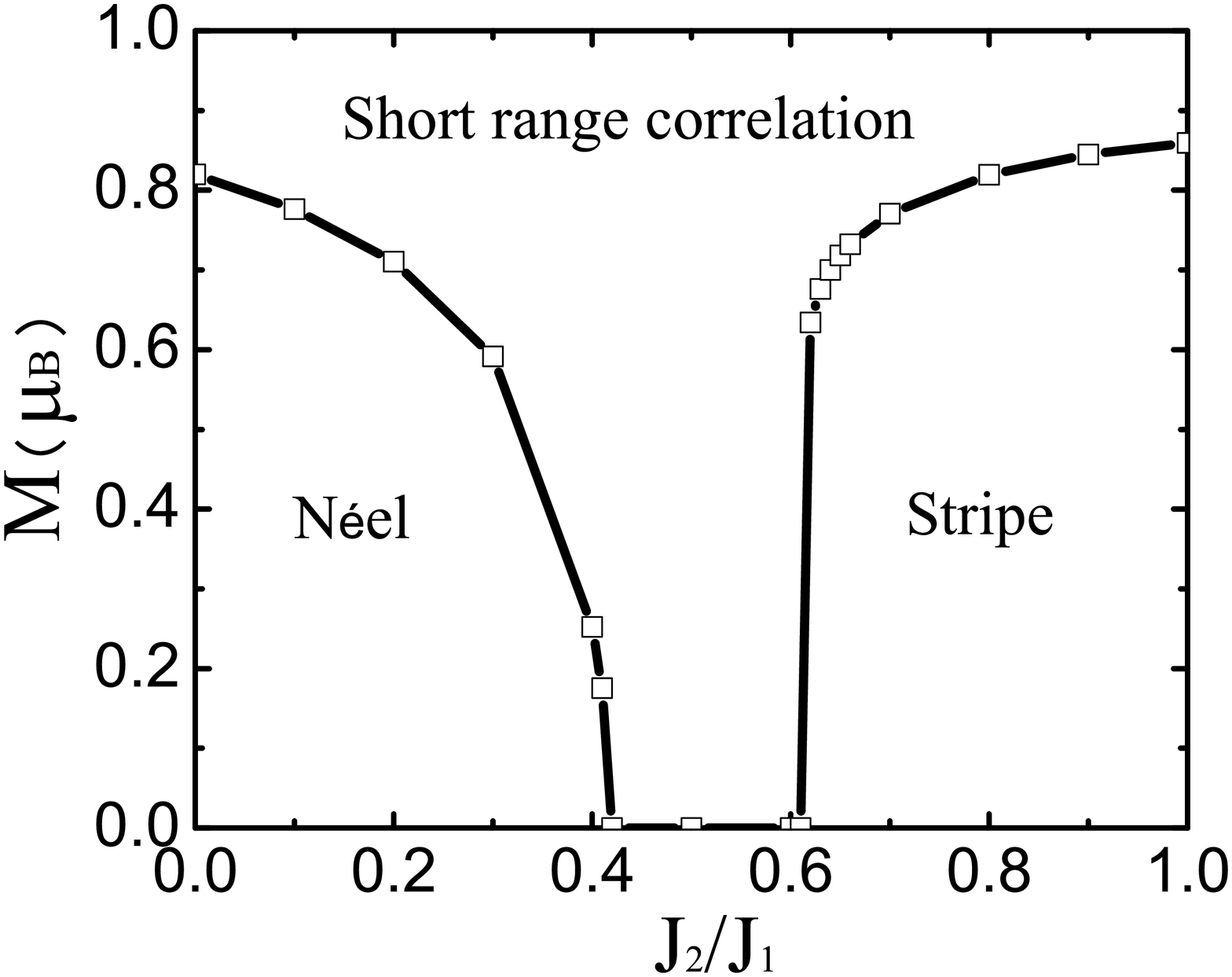}
\end{minipage}
\  \\
\caption{(a) Sketch of the square-lattice cluster adopted in our Cluster-SCF approach, and (b) phase diagram of $J_{1}$-$J_{2}$ spin-$\frac{1}{2}$ Heisenberg model obtained by our Cluster-SCF method.}
\label{Fig1}
\end{figure}

An effective $J_{1a}$-$J_{1b}$-$J_{2}$ Heisenberg model for iron-based compounds is described as,
\begin{equation}
\label{eq.1}
\begin{aligned}
H_{s}=&J_{1a}^{s}\sum_{\substack{<ij>_{a}}}\vec{s}_{i}\cdot\vec{s}_{j}
+J_{1b}^{s}\sum_{\substack{<ij>_{b}}}\vec{s}_{i}\cdot\vec{s}_{j}\\
&+J_{2}^{s}\sum_{\substack{<<ij>>}}\vec{s}_{i}\cdot\vec{s}_{j}+J_{c}^{s}\sum_{\substack{<ij>_{c}}}\vec{s}_{i}\cdot\vec{s}_{j}\\
&+J_{s}\sum_{\substack{i}}(s^{\alpha}_{i})^{2},
\end{aligned}
\end{equation}
where $J^{s}_{1a}$, $J^{s}_{1b}$ and $J^{s}_{2}$ are the nearest-neighbor (NN) and next-nearest-neighbor (NNN) magnetic exchange constants with spin $\vec{s}$ of Fe ions, respectively; $J_{c}^{s}$ is the interlayer coupling along the $c$-axis and $J_{s}$ is the single-ion anisotropy energy of Fe ions. Notice that $\alpha$ is given according to the experimental results, $\alpha$=$x$ ({\it i.e.} $a$ direction) for Fe spins in both EuFe$_{2}$As$_{2}$ and SmFeAsO compounds.

We consider the magnetic couplings in the rare-earth layer as
\begin{equation}
\label{eq.2}
\begin{aligned}
H_{S}=&J_{1}^{S}\sum_{\substack{<ij>}}\vec{S}_{i}\cdot\vec{S}_{j}
+J_{c}^{S}\sum_{\substack{<ij>_{c}}}\vec{S}_{i}\cdot\vec{S}_{j}\\
&+J_{S}\sum_{\substack{i}}(S^{\alpha}_{i})^{2},
\end{aligned}
\end{equation}
where $\vec{S}$ is the rare-earth spin, $J^{S}_{1}$ is the NN magnetic exchange constants of intralayer rare-earth spins, $J_{c}^{S}$ is the interlayer coupling along the $c$-axis, $J_{S}$ is the single-ion anisotropy energy of rare-earth ions, and $\alpha$=$x$ for Eu spins in EuFe$_{2}$As$_{2}$ and $\alpha$=$z$ ({\it i.e.} $c$ direction) for Sm spins in SmFeAsO. And we also consider the interlayer coupling between the rare-earth (R) and FeAs layers:
\begin{equation}
\label{eq.3}
\begin{aligned}
H_{sS}=&J_{c}^{sS}\sum_{\substack{<ij>_{c}}}\vec{s}_{i}\cdot\vec{S}_{j},
\end{aligned}
\end{equation}
where $J_{c}^{sS}$ is the coupling between R and Fe ions.

It is known that the external magnetic field is an effective tool to modulate the spin degree of freedom. In this paper we mainly consider two kinds of magnetic field, one ($B_{//}$) is applied along the direction of $a$ or $b$ axis of the magnetic unit cell, and another ($B_{\perp}$) is perpendicular to it ({\it i.e.} along the direction of $c$ axis). In general, the external magnetic field is described as
\begin{equation}
\label{eq.4}
\begin{aligned}
H_{B}=&-g\mu_{B}\mathbf{B}\sum_{\substack{i}}(s_{i}^{\alpha}+S_{i}^{\alpha})
\end{aligned}
\end{equation}
where $\alpha=x/z$ depends on the direction of the magnetic field, where $x$ ($z$) along $a$ ($c$) axis of the magnetic unit cell. Thus the total Hamiltonian of the system is $H=H_{s}+H_{S}+H_{sS}+H_{B}$.

In order to treat with the spin correlations and fluctuations including short-range ones accurately, we adopt the cluster self-consistent field (Cluster-SCF) method developed by us to solve this anisotropic Heisenberg model. The main idea of this method as follows: we divide the lattice into a central cluster plus surrounding spins, treat the magnetic interactions of the spins inside the cluster exactly, and the couplings of surrounding spins outside the cluster is treated as self-consistent "molecular" fields. The details could be found in Refs.~\cite{IJMPB21-691,JPCM21-026014,CPB18-4497}. Here we extend our method to deal with the Heisenberg model including the NNN interaction, such as $J_{1}$-$J_{2}$ and $J_{1a}$-$J_{1b}$-$J_{2}$ models, {\it etc.}. As an example to verify our approach, we calculated the phase diagram of $J_{1}$-$J_{2}$ Heisenberg model with spin $s=1/2$ on a square lattice, and the result is shown in Fig.~\ref{Fig1}. It is clearly shown that our phase diagram is good agreement with these obtained by other methods, such as series expansions~\cite{PRB54-3022} and coupled cluster~\cite{PRB78-214415} methods, demonstrating the effectiveness and validity of our Cluster-SCF method.

\begin{figure}[htbp]
\centering
\includegraphics[angle=0, width=0.4 \columnwidth]{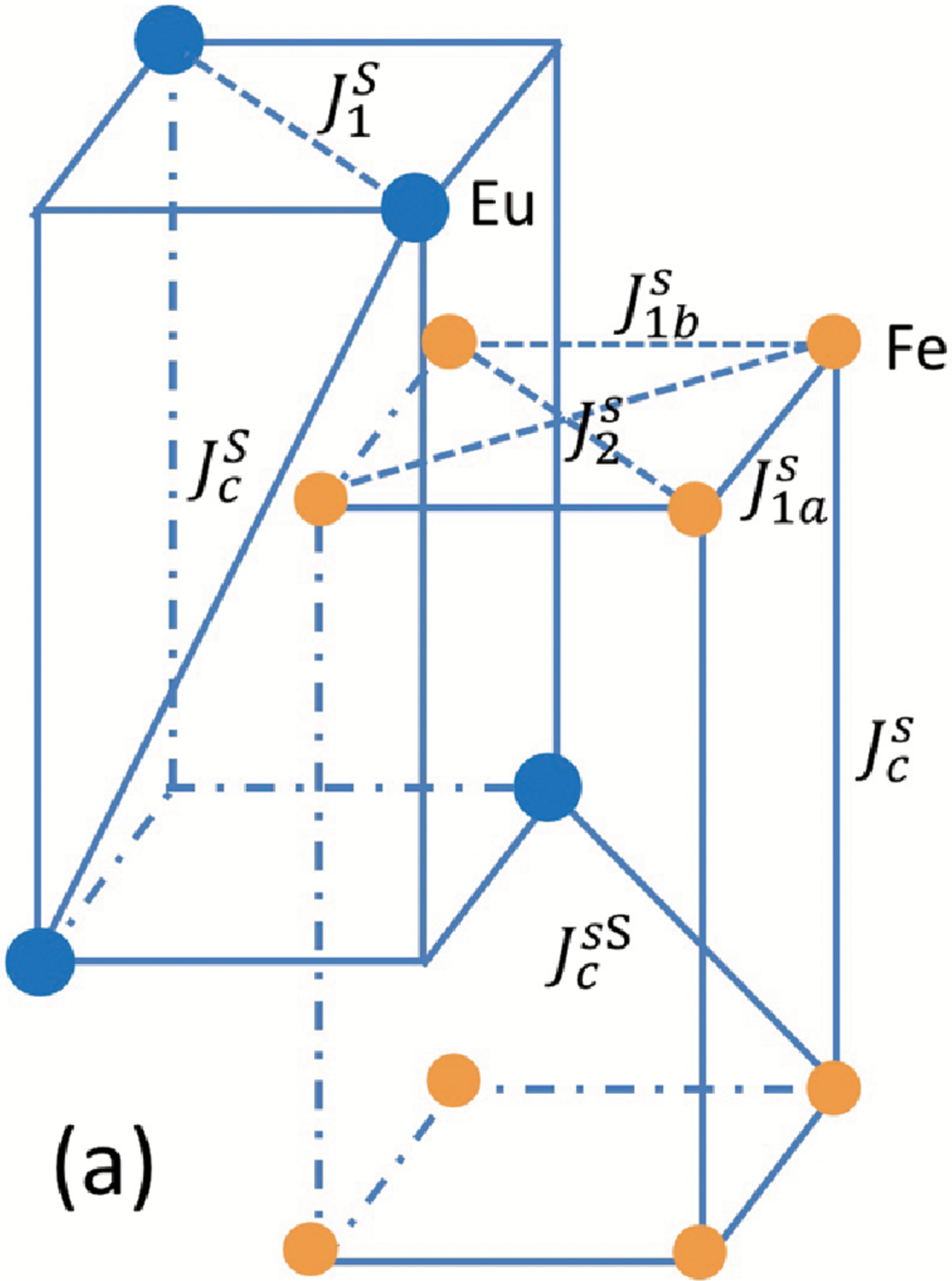}
\includegraphics[angle=0, width=0.4 \columnwidth]{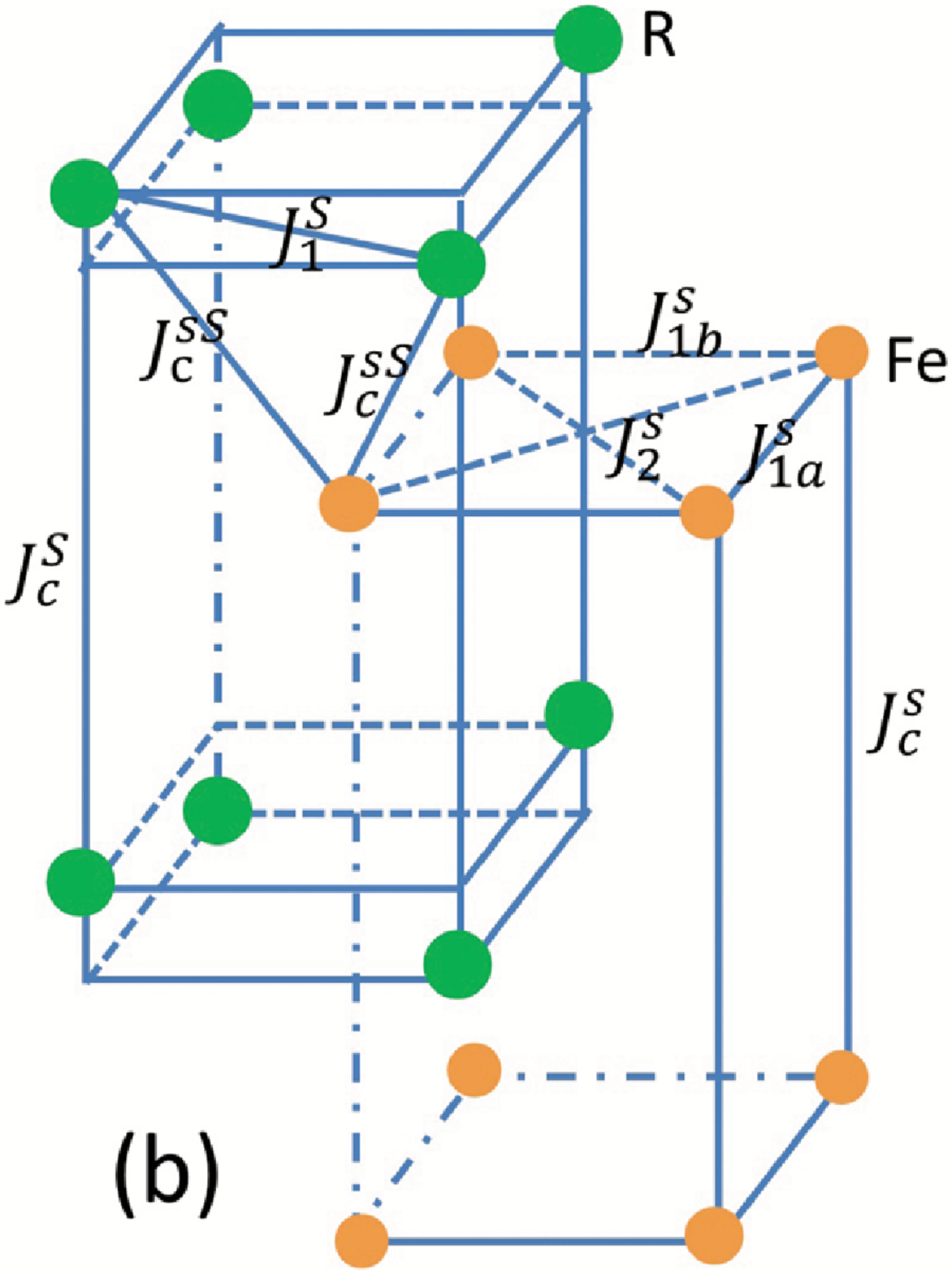}
\caption{Schematic clusters of (a) RFe$_{2}$As$_{2}$ (R=Eu) and (b) RFeAsO (R=Ce, Pr, Nd, Sm, Gd and Tb). All the magnetic exchange couplings are denoted.}
\label{Fig2}
\end{figure}

To solve this Heisenberg model, the magnetic exchange parameters should be given firstly. In order to compare the different rare-earth layers, we use the same parameters for Fe layers in both RFe$_{2}$As$_{2}$ and RFeAsO systems. Here we adopt the parameters for Fe layers with $J_{1a}^{s}$=59.9 meV, $J_{1b}^{s}$=$-$9.2 meV, $J_{2}^{s}$=13.6 meV and $J^{s}_{c}$=1.8 meV according to the inelastic neutron scattering experiments~\cite{nphys5-555,PRB84-054544}. And we estimate the single-ion anisotropy energy parameter $J_{s}$=0.07 meV. Note that we estimate the single-ion anisotropy energy parameters within spin-wave theory with a relationship $\Delta=2s\sqrt{2J_{s}(J_{1a}+2J_{2}+J_{c})}$, where $\Delta$ is spin gap~\cite{PRB87-140509,PRB78-220501} in this paper. For RFe$_{2}$As$_{2}$ case, considering N$\acute{e}$el transition temperature T$_{N}^{R}$=19 K for Eu layer in EuFe$_{2}$As$_{2}$~\cite{PRB78-052501,PRB78-052502}, we choose $J_{1}^{S}$=$-$0.8 meV, $J_{c}^{S}$=0.4 meV and $J_{S}$=0.2 meV. Due to the large magnetic moment $M_{Eu}$=6.9 $\mu_B$~\cite{PRB80-174424}, the spin $S=7/2$ for Eu ion is treated in a classical level. Meanwhile the spin $s=1$ for Fe ion is considered in a quantum level. The interlayer coupling between Eu and Fe ions $J^{sS}_{c}$=0.4 meV~\cite{PRB80-094524}. For RFeAsO (R=Ce, Pr, Nd, Sm, Gd and Tb) case, we choose $J_{1}^{S}$=4 meV, $J^{S}_{c}$=0.4 meV, $J_{S}$=0.2 meV and $J^{sS}_{c}$=0.4 meV~\cite{PRB80-094524}. In comparison with RFe$_{2}$As$_{2}$ case, the spin for R=Ce, Pr, Nd, Sm, Gd and Tb ions is also calculated with a classical value $S=1/2$~\cite{nphys8-709} due to a small magnetic moment ({\it e.g.} about 0.83 $\mu_B$ for CeFeAsO) for rear-earth ions in RFeAsO compounds~\cite{PRB80-094524}. The clusters and magnetic exchange couplings of RFe$_{2}$As$_{2}$ (R=Eu) and RFeAsO (R=Ce, Pr, Nd, Sm, Gd and Tb) compounds are shown in Fig.~\ref{Fig2}.

\section{Results and Discussions}

Utilizing the cluster-SCF method, we study the influence of magnetic field and the coupling between Fe and rare-earth spins on the magnetic ground states and magnetic phase transition both in RFe$_{2}$As$_{2}$ (R=Eu) and in RFeAsO (R=Ce, Pr, Nd, Sm, Gd and Tb) systems. In the following, we address the detailed results for these two systems for comparison.

\subsection{RFe$_{2}$As$_{2}$ case}

\begin{figure}[htbp]
\centering
\includegraphics[angle=0, width=0.75 \columnwidth]{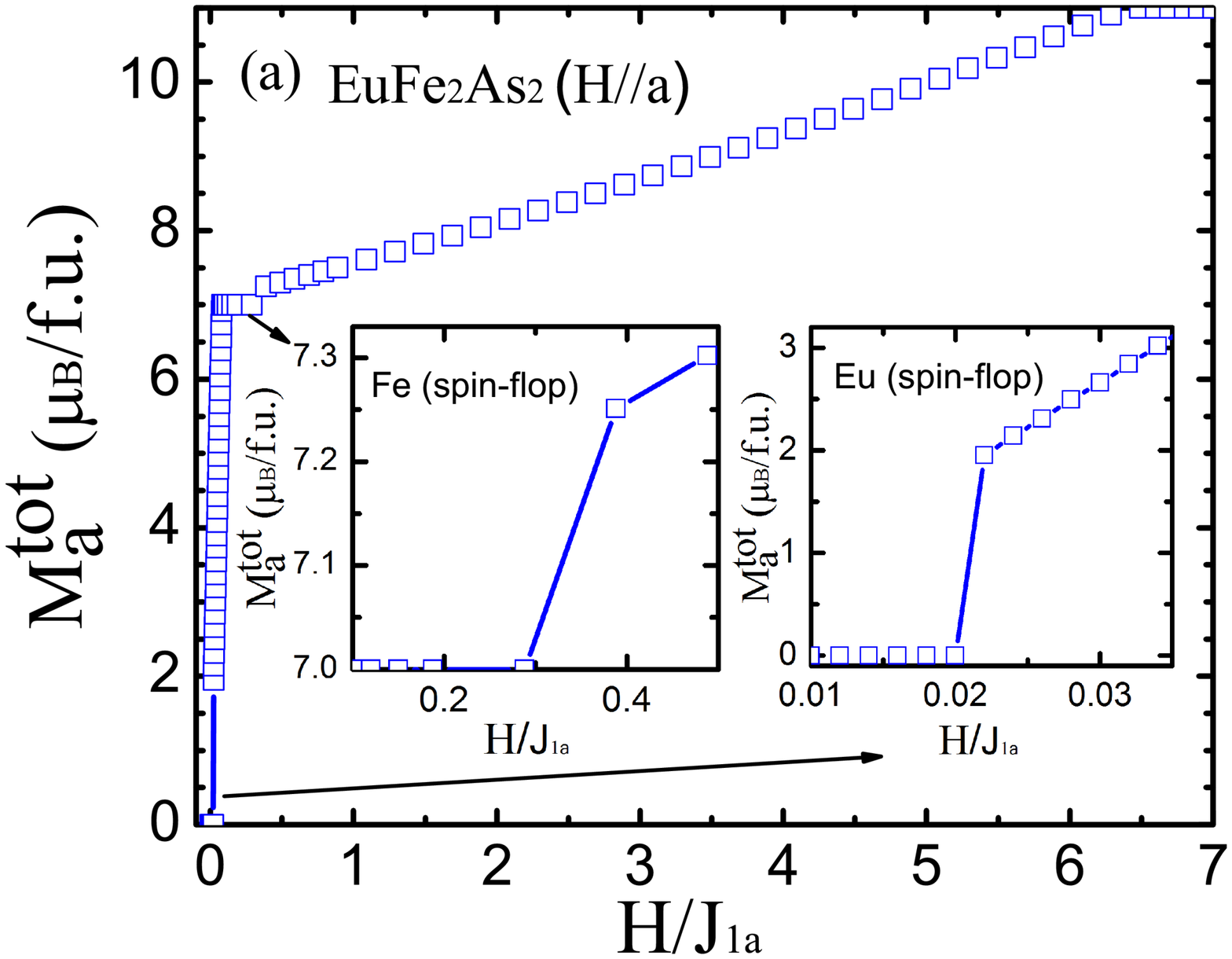}
\includegraphics[angle=0, width=0.75 \columnwidth]{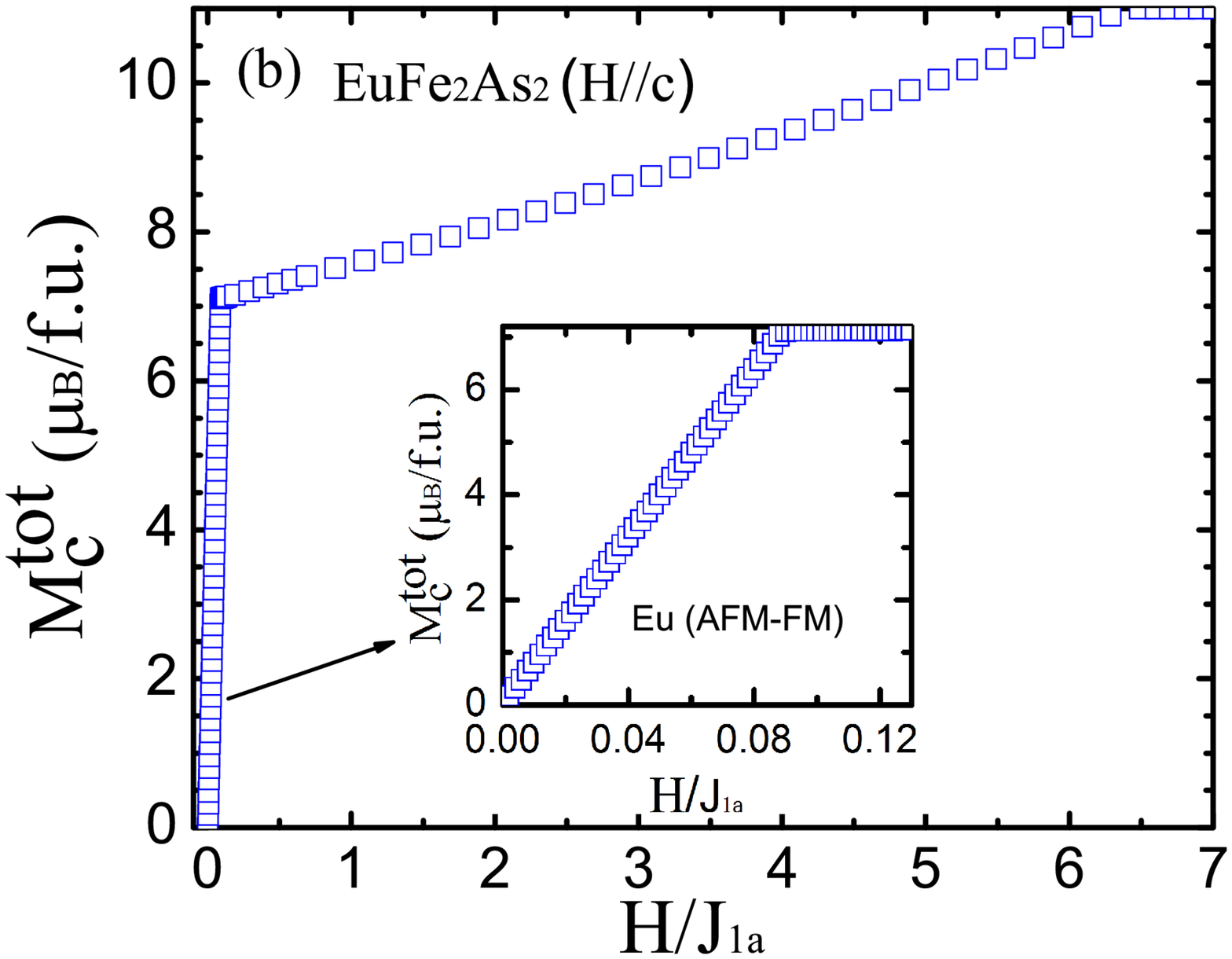}
\caption{Total magnetization dependence on magnetic field (a) $H//a$ and (b) $H//c$ in EuFe$_{2}$As$_{2}$. The left and right insets in (a) present the spin-flop transition, and the inset in (b) corresponds a successive AFM-FM transition.}
\label{Fig3}
\end{figure}

We find that in the stable magnetic ground state of RFe$_{2}$As$_{2}$ with R=Eu, the Fe spins in FeAs layer are SAFM with interlayer AFM; meanwhile, the Eu$^{2+}$ spins in the rare-earth layers are ferromagnetic (FM) with the interlayer AFM. Notice that the spin directions of both Eu$^{2+}$ and Fe$^{2+}$ ions align in the $a$-$b$ plane. According to the analysis of the symmetry of the magnetic structure, the influence of rare-earth ions on Fe ions is canceled in the absence of magnetic field within the present mean field approximation. Once the magnetic field is applied, the magnetic rare-earth ions would contribute an effective molecular field on Fe spins. Our results show that when a magnetic field ($H//a$ or $H//c$) is applied, the system undergoes a series of complicated magnetic phase transitions. As shown in Fig.~\ref{Fig3}(a) and (b), the magnetic field dependence of total magnetization per formula ($M_{a}^{tot}$ and $M_{c}^{tot}$) globally displays similar behavior for $H//a$ and $H//c$. With the increase of the magnetic field, $M_{a}^{tot}$ and $M_{c}^{tot}$ first undergo a lift steeply, and then almost linearly increase to a saturated value.

However, the detail of $M_{a}^{tot}$ displays a lot of difference from that of $M_{c}^{tot}$. The total magnetization $M_{a}^{tot}$ of EuFe$_{2}$As$_{2}$ displays two sharp phase transitions under the parallel magnetic filed $H//a$: the first one corresponds to the spin flop transition of Eu$^{2+}$ spins and the second one to that of Fe$^{2+}$ spins, as seen in the two insets of Fig.~\ref{Fig3}(a). The critical magnetic field of the spin-flop transition for Eu$^{2+}$ ions, $H_{c}$(Eu), is about 0.02$J_{1a}$ ($\sim$9 T). In fact, due to the low critical magnetic field, the spin-flop transition for Eu$^{2+}$ ions is observed in the experiments~\cite{NJP11-025007,PRB81-220406,JLTP159-601}. While for a perpendicular magnetic field ($H//c$), no spin-flop transition is observed, only a successive AFM-FM transition corresponds a canted magnetism of rare-earth ions. When the magnetic field becomes strong enough, the total magnetization ferromagnetically saturates in both cases.

\begin{figure}[htbp]
\centering
\includegraphics[angle=0, width=0.45 \columnwidth]{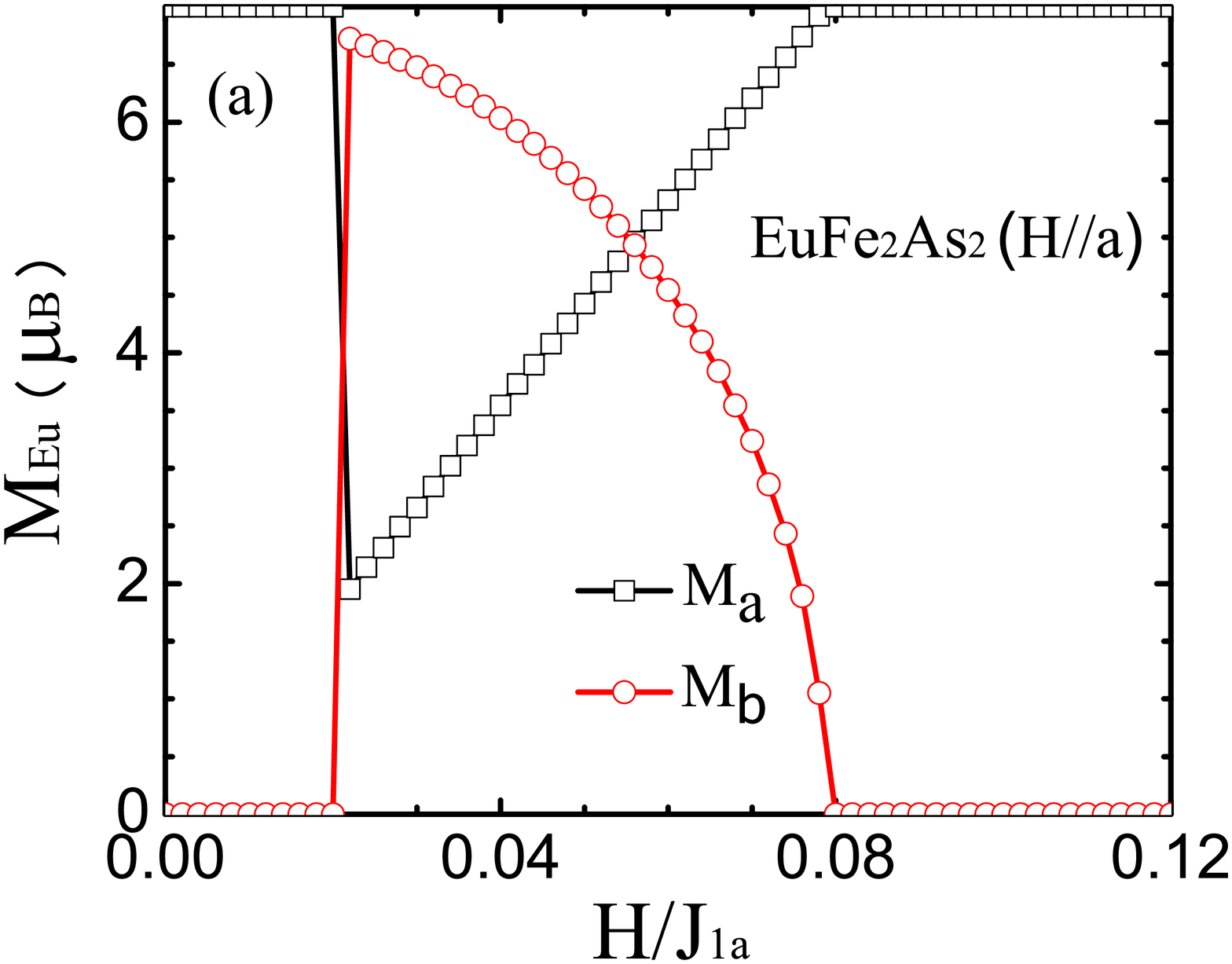}
\includegraphics[angle=0, width=0.45 \columnwidth]{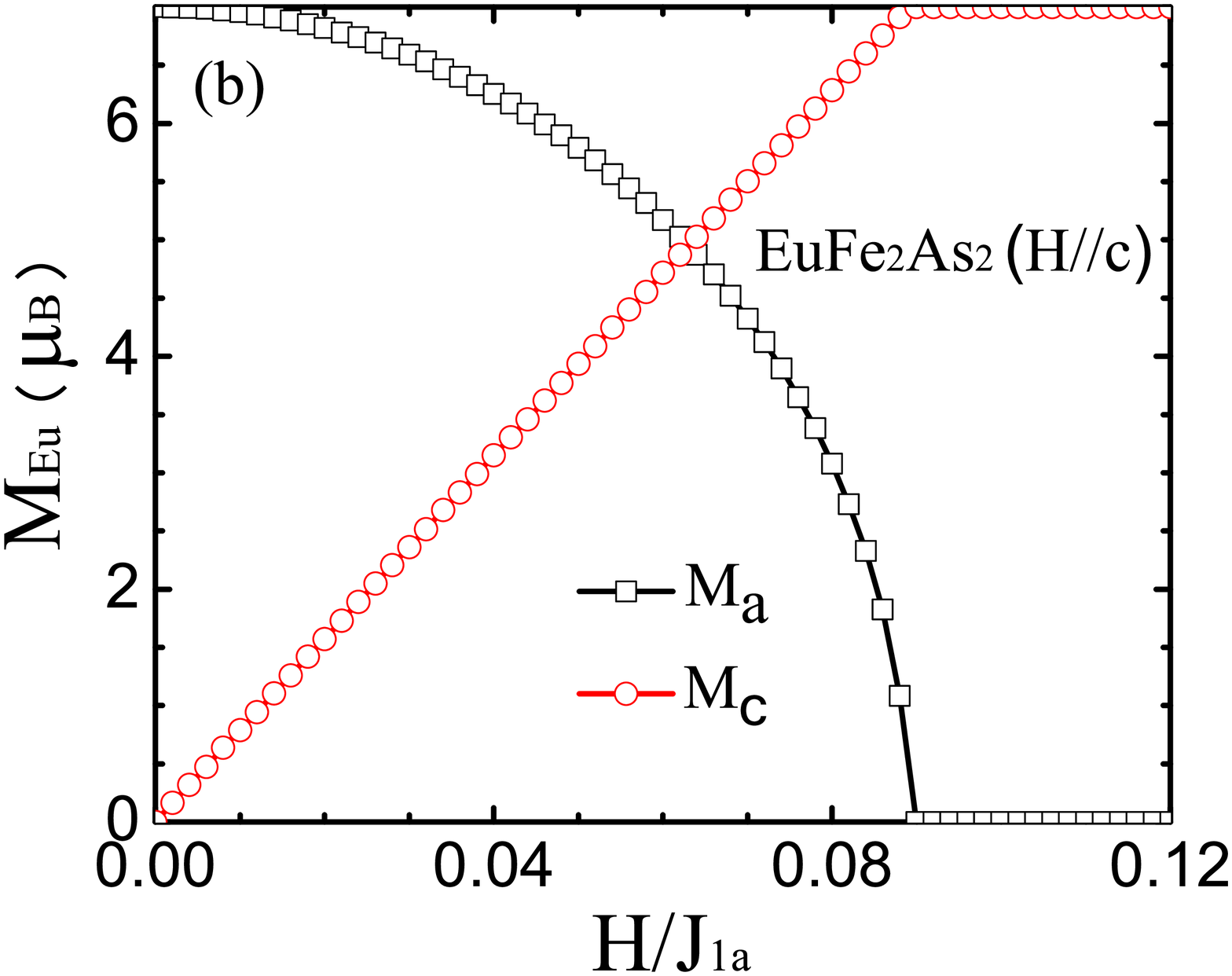}
\includegraphics[angle=0, width=0.45 \columnwidth]{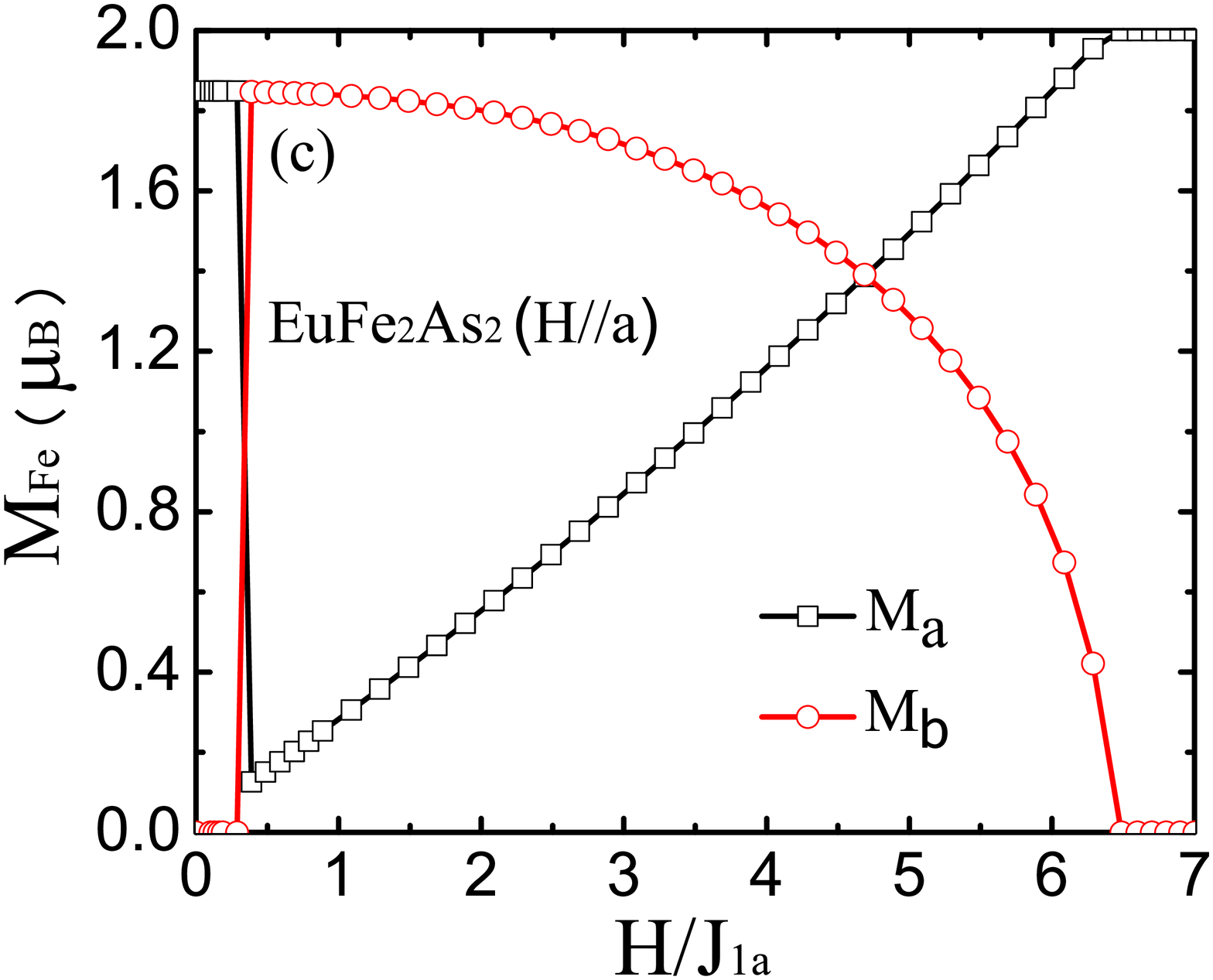}
\includegraphics[angle=0, width=0.45 \columnwidth]{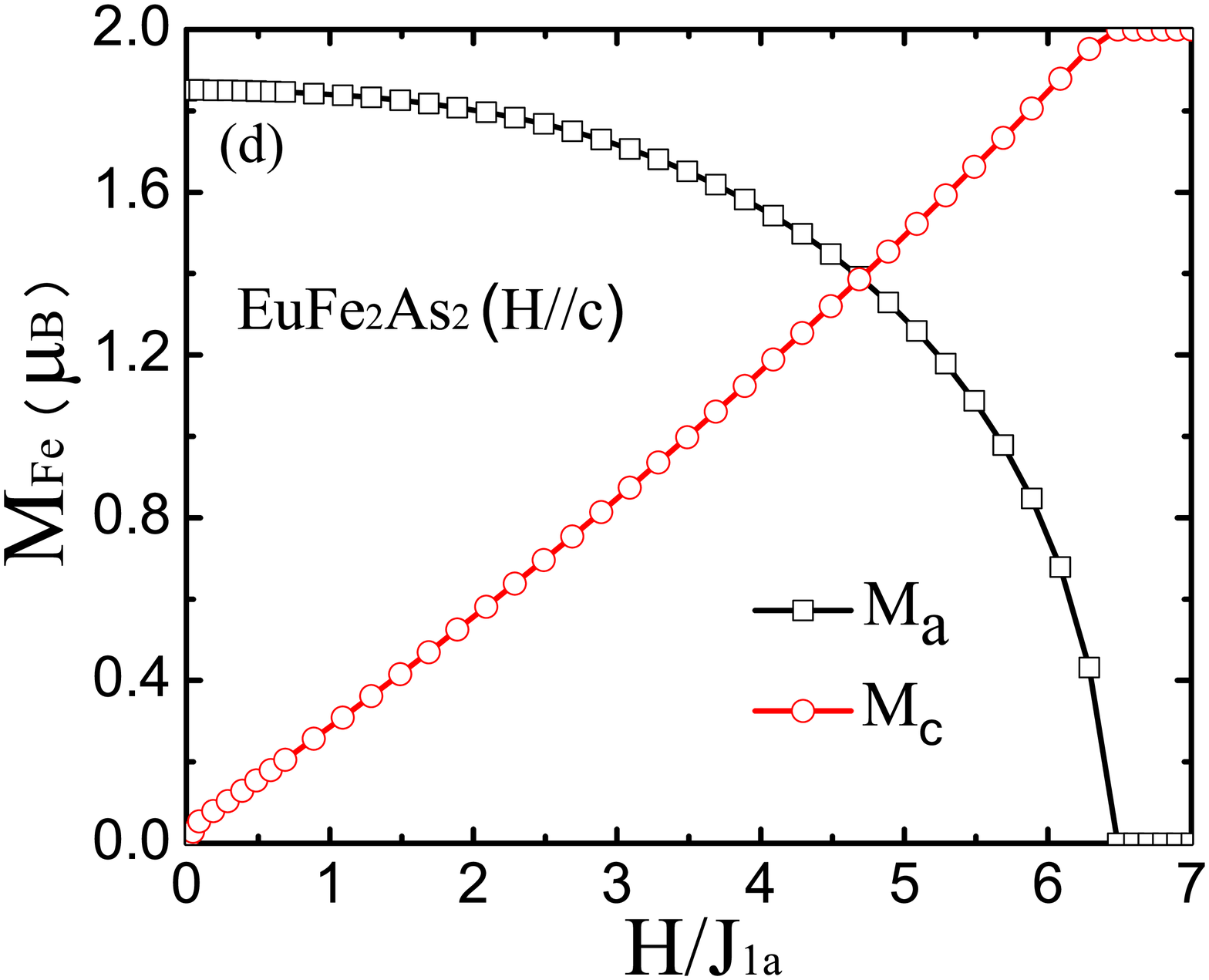}
\caption{Dependence of magnetization components, $M_{a}$ and $M_{b}$, of Eu$^{2+}$ ions and of Fe$^{2+}$ ions on magnetic field (a) and (c) $H//a$ and (b) and (d) $H//c$ in EuFe$_{2}$As$_{2}$.}
\label{Fig4}
\end{figure}

The atom-resolved magnetization contributed from Eu$^{2+}$ and Fe$^{2+}$ ions are plotted for $H//a$ and $H//c$ in Fig.~\ref{Fig4}. From Fig.~\ref{Fig4}(a), one can see that when applied parallel magnetic field ($H//a$) becomes larger than a critical field $H_{c}^{R}$, the magnetization components $M_{a}$ and $M_{b}$ of Eu ions suffer a sharp change, corresponding to a spin flop transition. With the further increase of $H$, the magnetic Eu ions enter into a canted phase, until into a saturated FM phase. In contrast, for a perpendicular magnetic field ($H//c$), once the magnetic field is applied, the system gradually transits to a canted phase, as seen in Fig.~\ref{Fig4}(b). One notices that at $H_{c}^{R}$, there is no influence of Fe ions on the rare-earth ions due to the cancellation of the AFM molecular fields within the present mean field approximation.
As a comparison with rare-earth 4$f$ spins, the magnetic phase transitions of Fe-3$d$ magnetism are similar, but occurs at a relatively large magnetic field. From Fig.~\ref{Fig4}(c) with $H//$a, it is clearly found that the critical parallel magnetic field of spin-flop for Fe$^{2+}$ ion ($H_{c}$(Fe)) is about 0.288$J_{1a}$ ($\sim$130 T), which is not observed experimentally due to its large $H_{c}$(Fe). Thus Fe-spin flop phenomenon may only be observed experimentally in a strong pulsed magnetic field. In fact, the effect of R ions on Fe ions is equivalent to an effective molecular field, about 0.11$J_{1a}$ at $H_{c}^{Fe}$. This shows that the 4$f$-magnetism could considerably affect the Fe-3$d$ magnetic properties under magnetic field.

\begin{figure}[htbp]
\centering
\includegraphics[angle=0, width=0.95 \columnwidth]{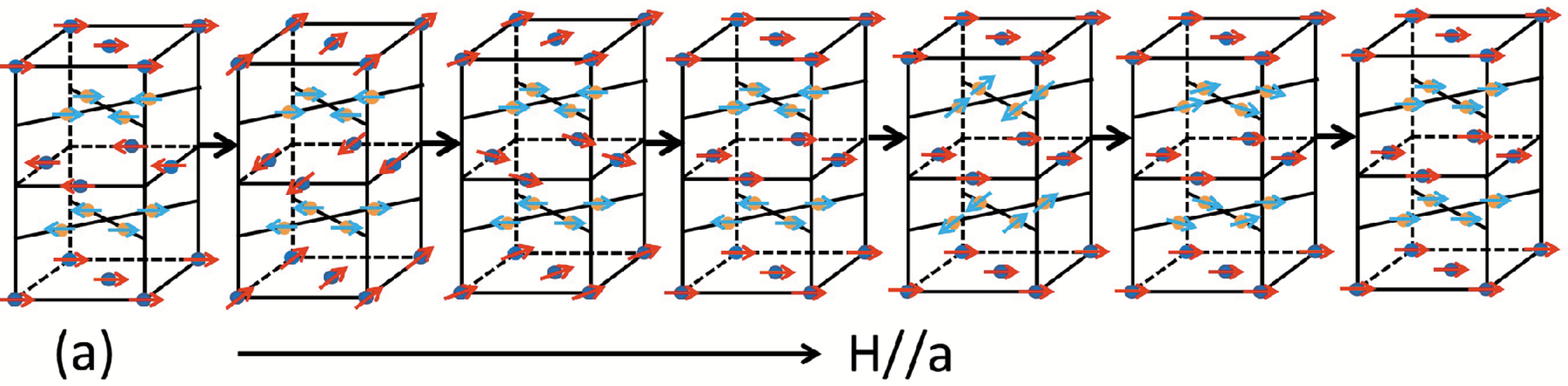}
\includegraphics[angle=0, width=0.7 \columnwidth]{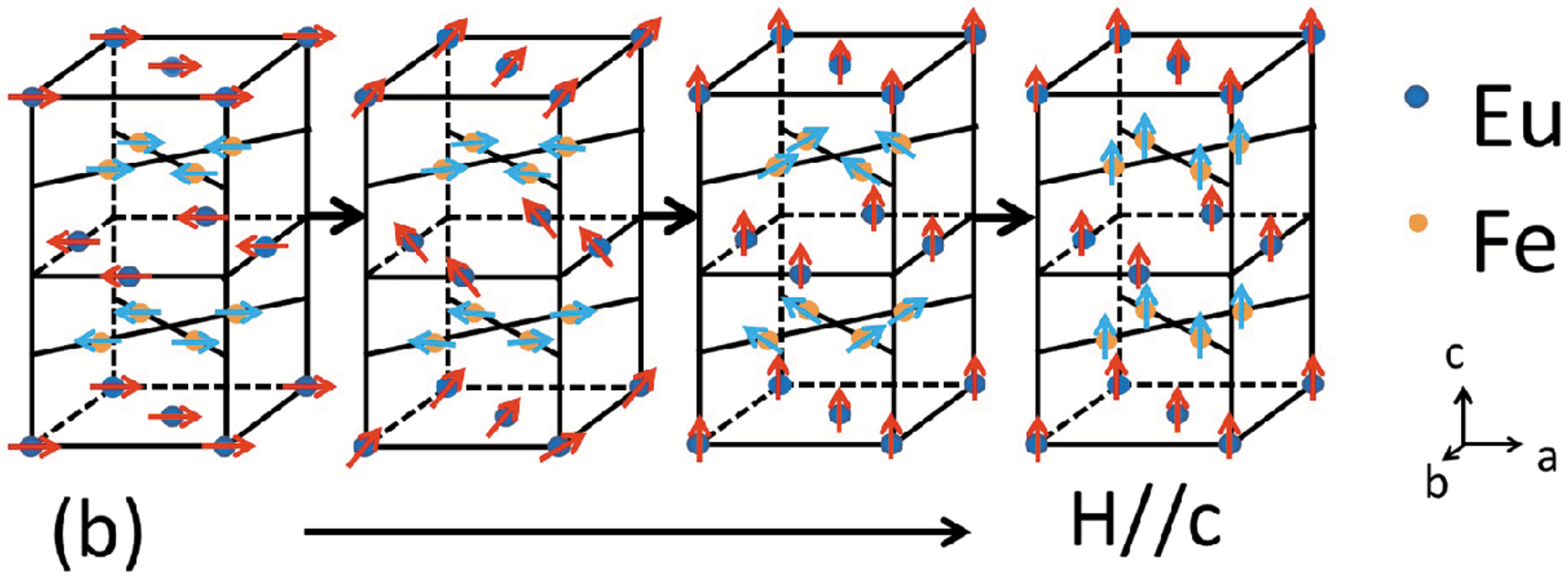}
\caption{Evolution of magnetic structure on the magnetic field $H//a$ (a) and $H//c$ (b) in EuFe$_{2}$As$_{2}$. The red and blue arrows represent the spins of Eu and Fe, respectively. The long arrow indicates the increasing direction of magnetic field strength.}
\label{Fig5}
\end{figure}

To further uncover the influence of magnetic field on the magnetic structures in EuFe$_{2}$As$_{2}$, we display the evolution of magnetic structure on the magnetic field $H$ in Fig.~\ref{Fig5}, where we denote the magnetic configuration of each stage in such a form: R(intralayer-interlayer coupling between R-R ions)-RFe(interlayer coupling between R-Fe ions)-Fe(intralayer-interlayer coupling between Fe-Fe ions). With the increasing magnetic field, the system undergoes the following configurations: under parallel field $H//a$, the system evolves from  (1) Eu(FM-AF)-AF-Fe(SAFM-AF), (2) Eu(spin-flop-FM-AF)-AF-Fe(SAFM-AF), (3) Eu(FM-canted)-AF-Fe(SAFM-AF), (4) Eu(FM-FM)-AF-Fe(SAFM-AF), (5) Eu(FM-FM)-AF-Fe(spin-flop-SAFM-AF), (6) Eu(FM-FM)-AF-Fe(canted-canted), to (7) Eu(FM-FM)-F-Fe(FM-FM), as seen in Fig.~\ref{Fig5}(a), from which one observes that the spin-flop transitions of Eu spins and Fe spins occur at the second and fifth stage, respectively; and under perpendicular field $H//c$, the system evolves from (1) Eu(FM-AF)-AF-Fe(SAFM-AF), (2) Eu(FM-canted)-AF-Fe(canted-canted), (3) Eu(FM-FM)-AF-Fe(canted-canted), to (4) Eu(FM-FM)-F-Fe(FM-FM), as seen in Fig.~\ref{Fig5}(b).

\begin{figure}[htbp]
\centering
\includegraphics[angle=0, width=0.75 \columnwidth]{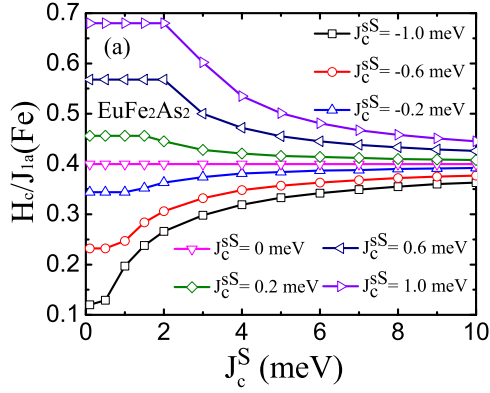}
\caption{Critical magnetic field of the spin-flop transition ($H_{c}$(Fe)) for Fe$^{2+}$ ion depends on the interlayer coupling $J_{c}^{S}$ in EuFe$_{2}$As$_{2}$ with magnetic field $H//a$.}
\label{Fig6}
\end{figure}

In simple systems, it is known that the critical magnetic field of the spin-flop transition is proportional to the single-ion anisotropy energy. In the present complicated magnetic systems, however, due to the interlayer coupling between Eu$^{2+}$ and Fe$^{2+}$ ions, the variation of the critical magnetic field is very complex, as seen in Fig.~\ref{Fig6}. It shows that the FM interlayer coupling between rare-earth and Fe ions ($J^{sS}_{c}<$0) favors the occurrence of the Fe-spin flop transition; however, the AFM interlayer coupling ($J^{sS}_{c}>$0) is unfavorable of the transition. Furthermore, the interlayer coupling between rare-earth ions, $J_{c}^{S}$, also strongly affects the spin-flop transition; for the FM (AFM) interlayer coupling between rare-earth and Fe ions, $J_{c}^{S}$, {\it i.e.} the AFM coupling between magnetic rare-earth ions, unfavors (favors) the occurrence of the Fe spin-flop transition. In addition, the intralayer coupling of rare-earth ions, $J_{1}^{S}$, does not affect this transition.

\subsection{RFeAsO case}

\begin{figure}[htbp]
\centering
\includegraphics[angle=0, width=0.75 \columnwidth]{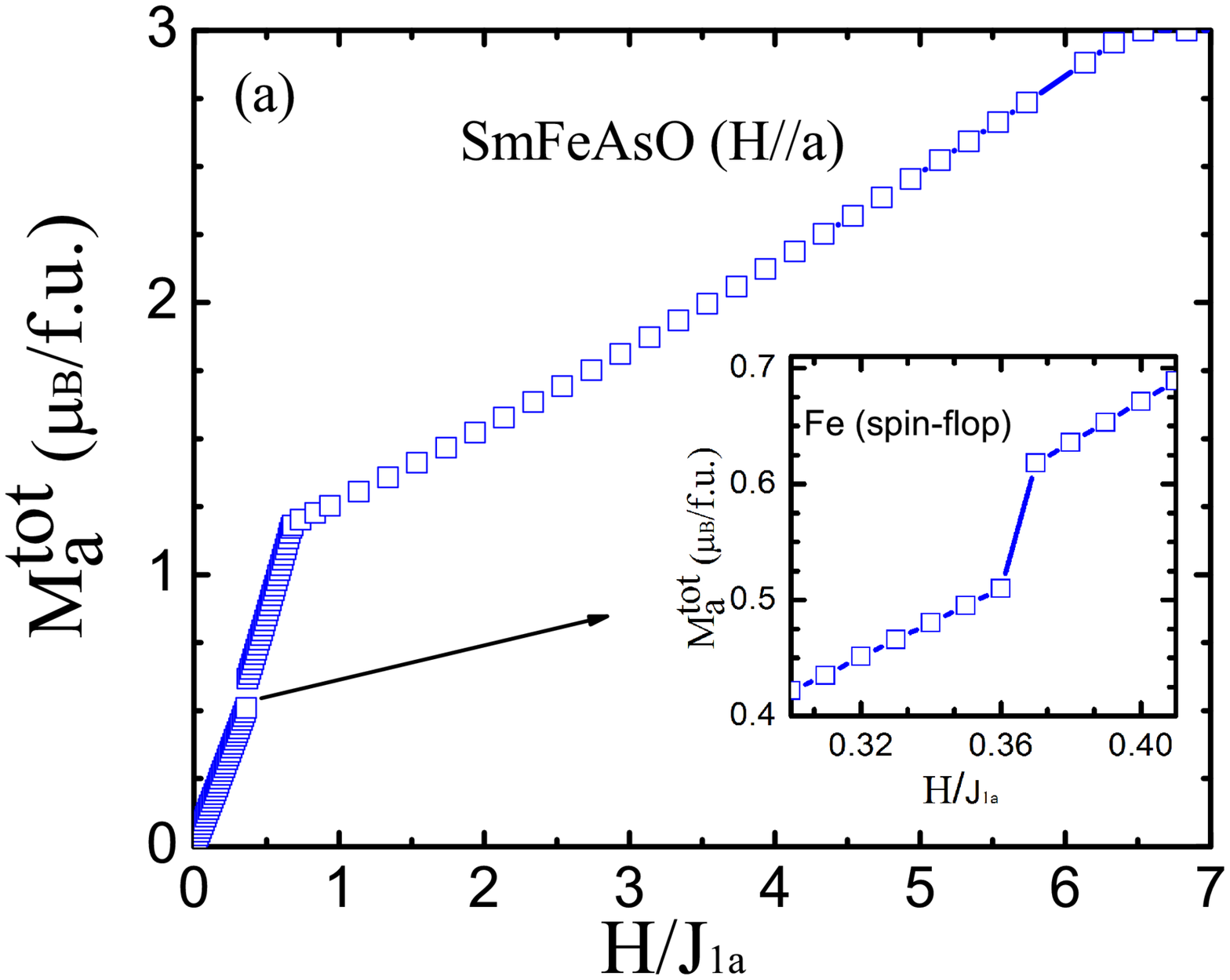}
\includegraphics[angle=0, width=0.75 \columnwidth]{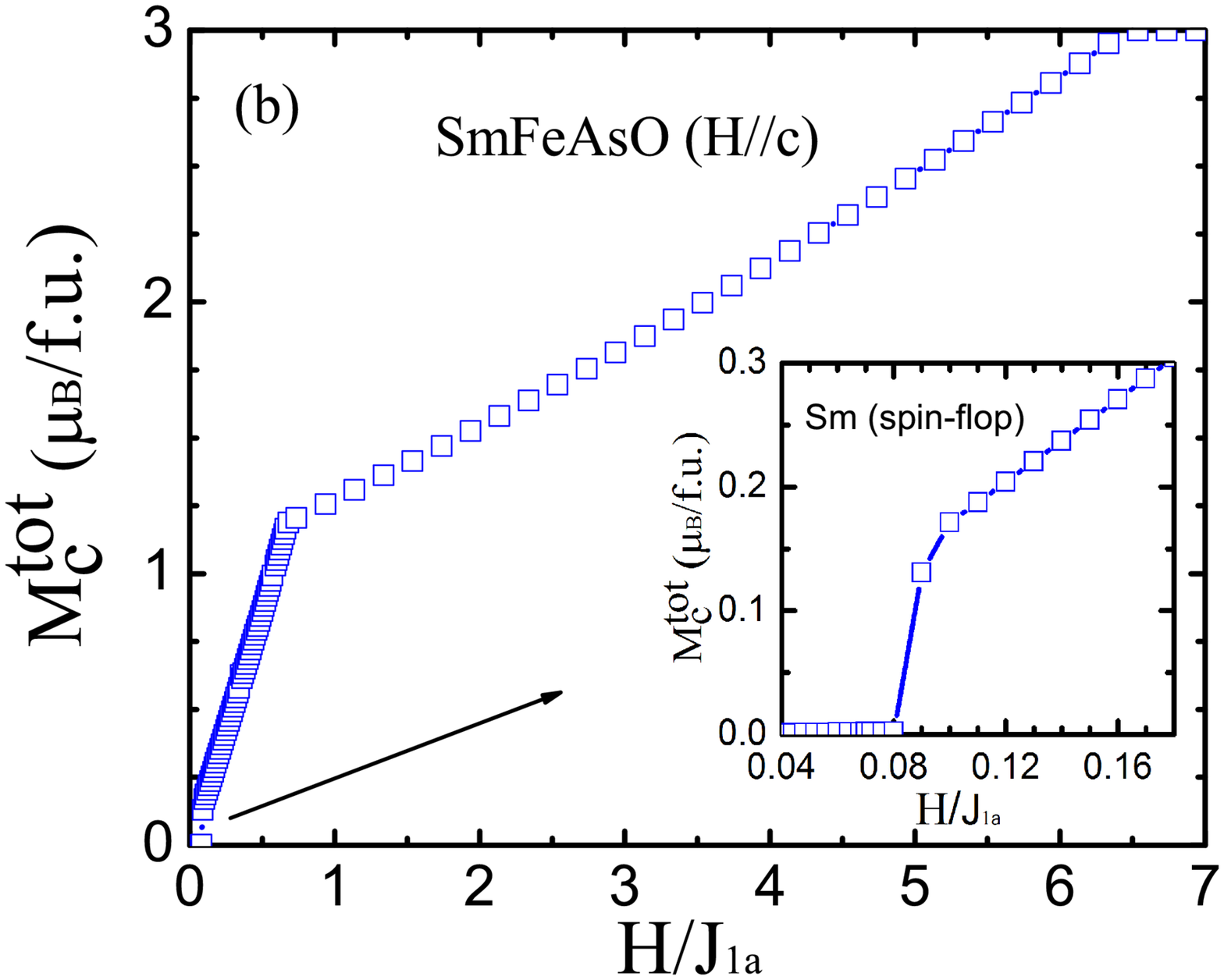}
\caption{Dependence of total magnetization on magnetic field (a) $H//a$ and (b) $H//c$ of SmFeAsO.}
\label{Fig7}
\end{figure}

Actually, the magnetism of 4$f$ electrons in RFeAsO (R=Ce, Pr, Nd, Sm Gd and Tb) is very different from that in RFe$_{2}$As$_{2}$. In order to investigate the magnetic interplay between 4$f$ and 3$d$ electrons, we also calculate the magnetic properties of RFeAsO (R=Sm). For RFeAsO (R=Sm), the calculated magnetic ground state of the rare-earth layer is N$\acute{e}$el-AFM (NAFM) with interlayer AFM, and that of Fe layer is SAFM with an AFM interlayer Fe spins. In the present case, the spin directions of Sm$^{2+}$ and Fe$^{2+}$ ions align along the $c$ axis and in the $a$-$b$ plane, respectively. As a consequence, when a magnetic field ($H//a$ or $H//c$) is applied, the system undergoes a complex magnetic phase transition and is different from EuFe$_{2}$As$_{2}$. As shown in Fig.~\ref{Fig7}, the total magnetization of SmFeAsO displays only one spin-flop transition under the parallel magnetic filed ($H//a$) in the FeAs layer. While for the perpendicular magnetic filed ($H//c$), the spin flop transition occurs in the rare earth layers. This is very different from the EuFe$_{2}$As$_{2}$ shown above.

\begin{figure}[htbp]
\centering
\includegraphics[angle=0, width=0.45 \columnwidth]{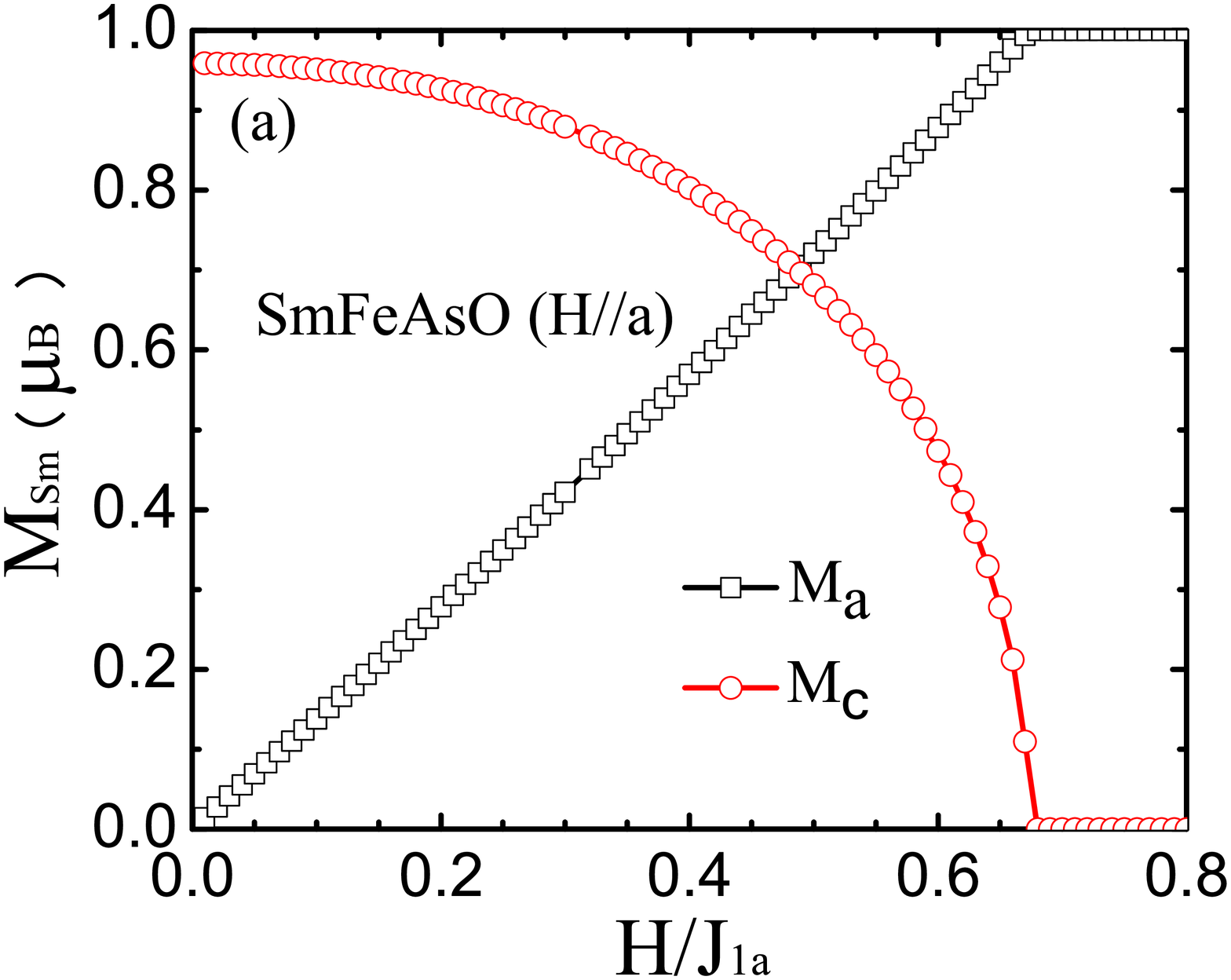}
\includegraphics[angle=0, width=0.45 \columnwidth]{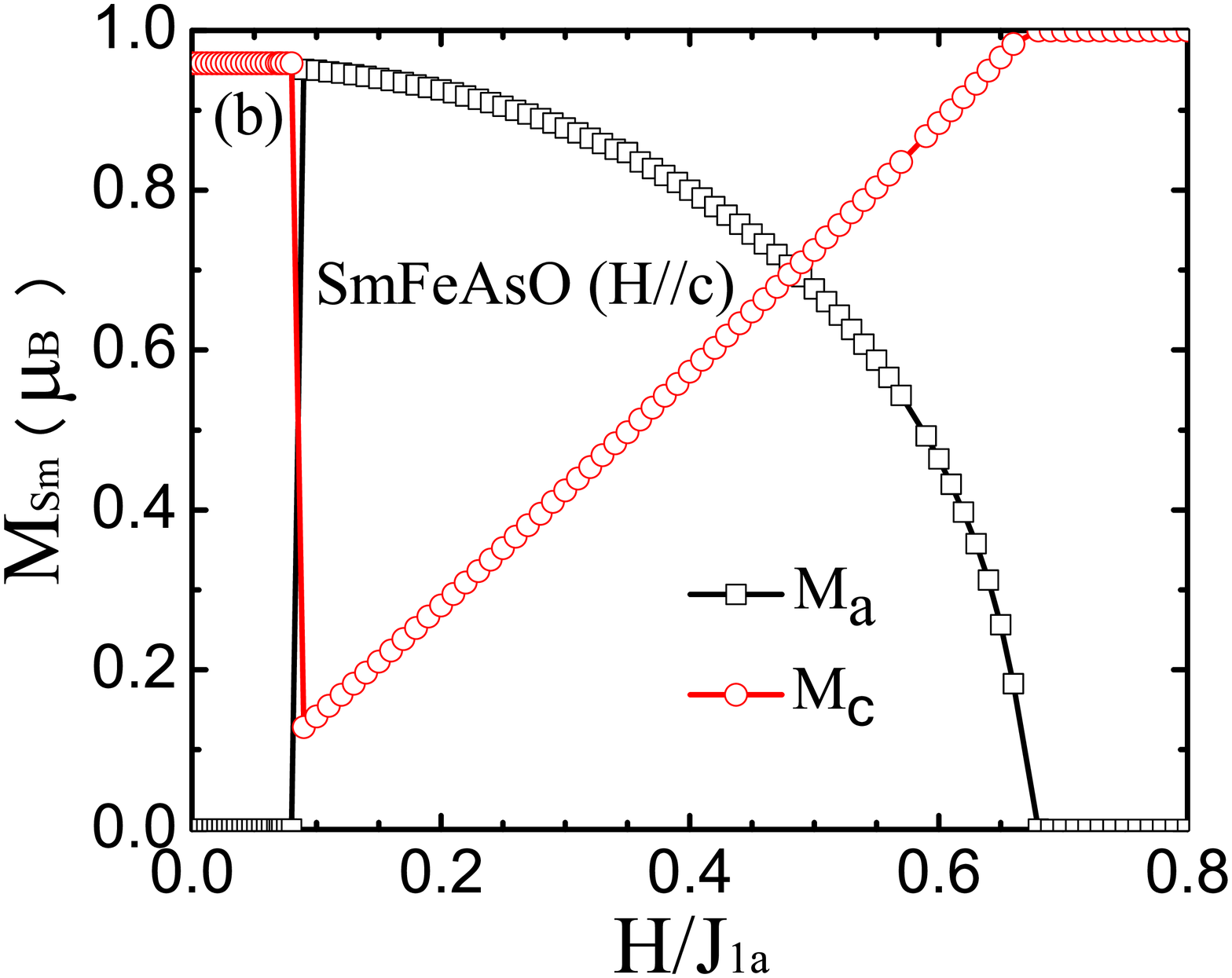}
\caption{Dependence of sublattice magnetization components($M_{a}$ and $M_{b}$) of Sm$^{2+}$ ion on magnetic field (a) $H//a$ and (b) $H//c$ in SmFeAsO.}
\label{Fig8}
\end{figure}

The atom-resolved magnetizations of Sm$^{2+}$ and Fe$^{2+}$ ions in SmFeAsO are also plotted in Fig.~\ref{Fig8}. From Fig.~\ref{Fig8}(b), it is obviously found that the critical magnetic field of spin-flop for Sm$^{2+}$ ions ($H_{c}$(Sm)) is about 0.08$J_{1a}$ ($\sim$36 T), considerably larger than the critical value in EuFe$_{2}$As$_{2}$. In fact, this spin-flop transition of Sm$^{2+}$ ions was observed in a strong pulsed magnetic field in a recent experiment~\cite{PRB83-134503}. In contrast to Eu ions in 122 system, the influence of the molecular field of Fe spins on Sm spins is significant due to different magnetic polarized axis of Fe and Sm ions, which contributes to a weak effective field about 0.0007$J_{1a}$ on Sm spins.
Meanwhile, we find that the critical magnetic field of spin-flop for Fe$^{2+}$ ions ($H_{c}$(Fe)) is relatively large, about 0.364$J_{1a}$ ($\sim$164 T). This critical value is slightly larger than that of EuFe$_{2}$As$_{2}$. The reason is that the magnetic field should firstly overcome the intralayer AFM coupling between Sm ions before the spin-flop transition occurs. And the effective field contributed from Sm ions on Fe ions is about 0.035$J_{1a}$.

\begin{figure}[htbp]
\centering
\includegraphics[angle=0, width=0.95 \columnwidth]{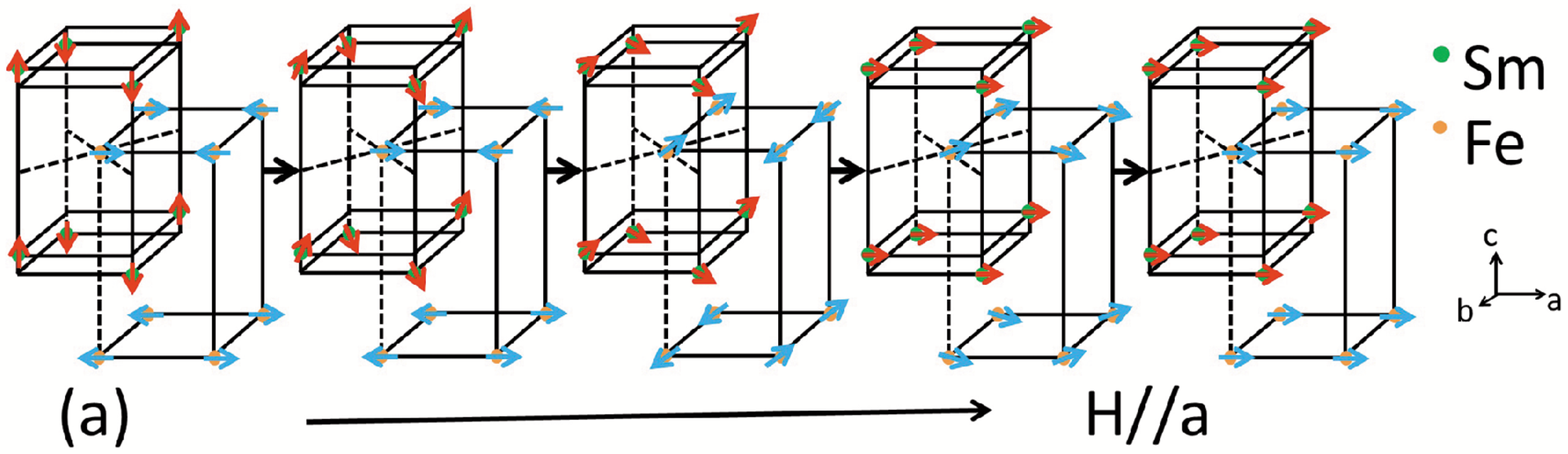}
\includegraphics[angle=0, width=0.95 \columnwidth]{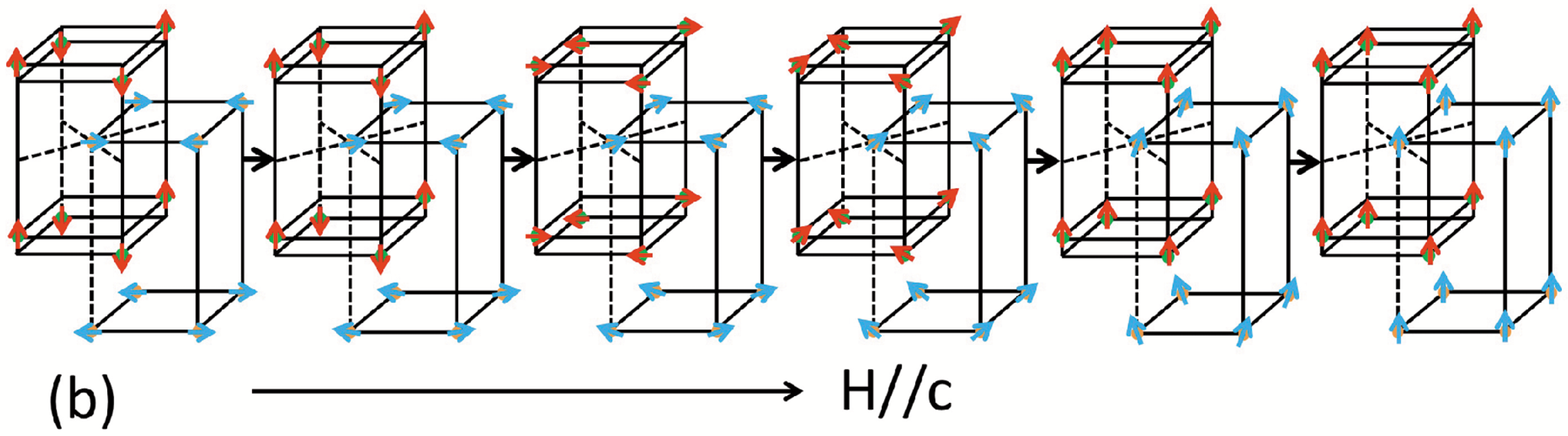}
\caption{Evolution of magnetic structure on applied magnetic field $H//a$ (a) and $H//c$ (b) in SmFeAsO.}
\label{Fig9}
\end{figure}

The evolution of magnetic structure in SmFeAsO on applied magnetic field $H$ is displayed in Fig.~\ref{Fig9}. With the same definition to last section, one finds that with the increase of the magnetic field, the system undergoes the following stages: under $H//a$ case, the system undergoes from (1) R(NAFM-AF)-AF-Fe(SAFM-AF), (2) R(canted-canted)-AF-Fe(SAFM-AF), (3) R(canted-canted)-AF-Fe(spin-flop-SAFM-AF), (4) R(FM-FM)-AF-Fe(canted-canted), to (5) R(FM-FM)-F-Fe(FM-FM); and under $H//c$ case, the system undergoes from (1) R(NAFM-AF)-AF-Fe(SAFM-AF), (2) R(NAFM-AF)-AF-Fe(canted-canted), (3) R(spin-flop-NAFM-AF)-AF-Fe(canted-canted), (4) R(canted-canted)-AF-Fe(canted-canted), (5) R(FM-FM)-AF-Fe(canted-canted), to (6) R(FM-FM)-F-Fe(FM-FM), obviously different from these in EuFe$_{2}$As$_{2}$ compounds.

\begin{figure}[htbp]
\centering
\includegraphics[angle=0, width=0.75 \columnwidth]{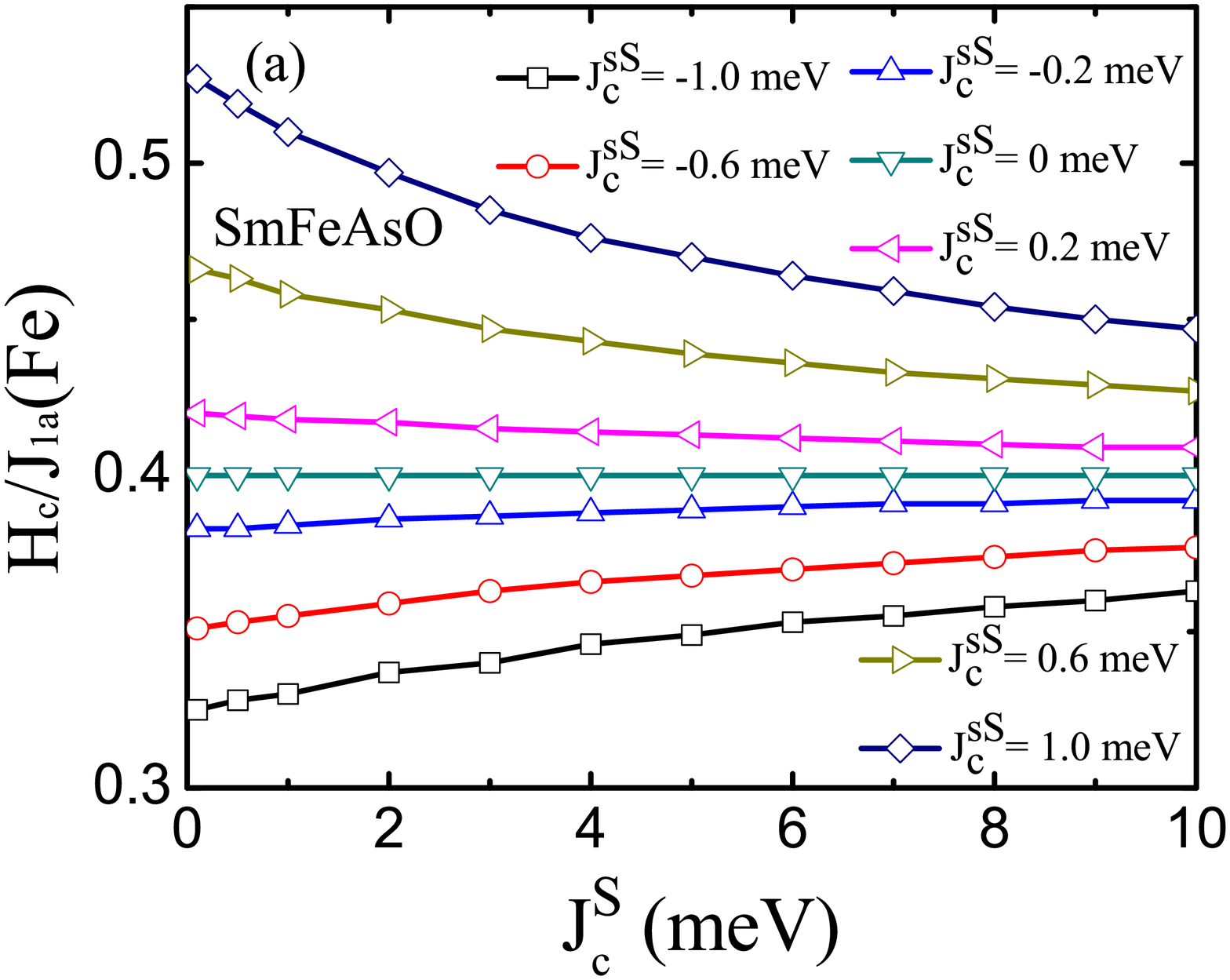}
\includegraphics[angle=0, width=0.75 \columnwidth]{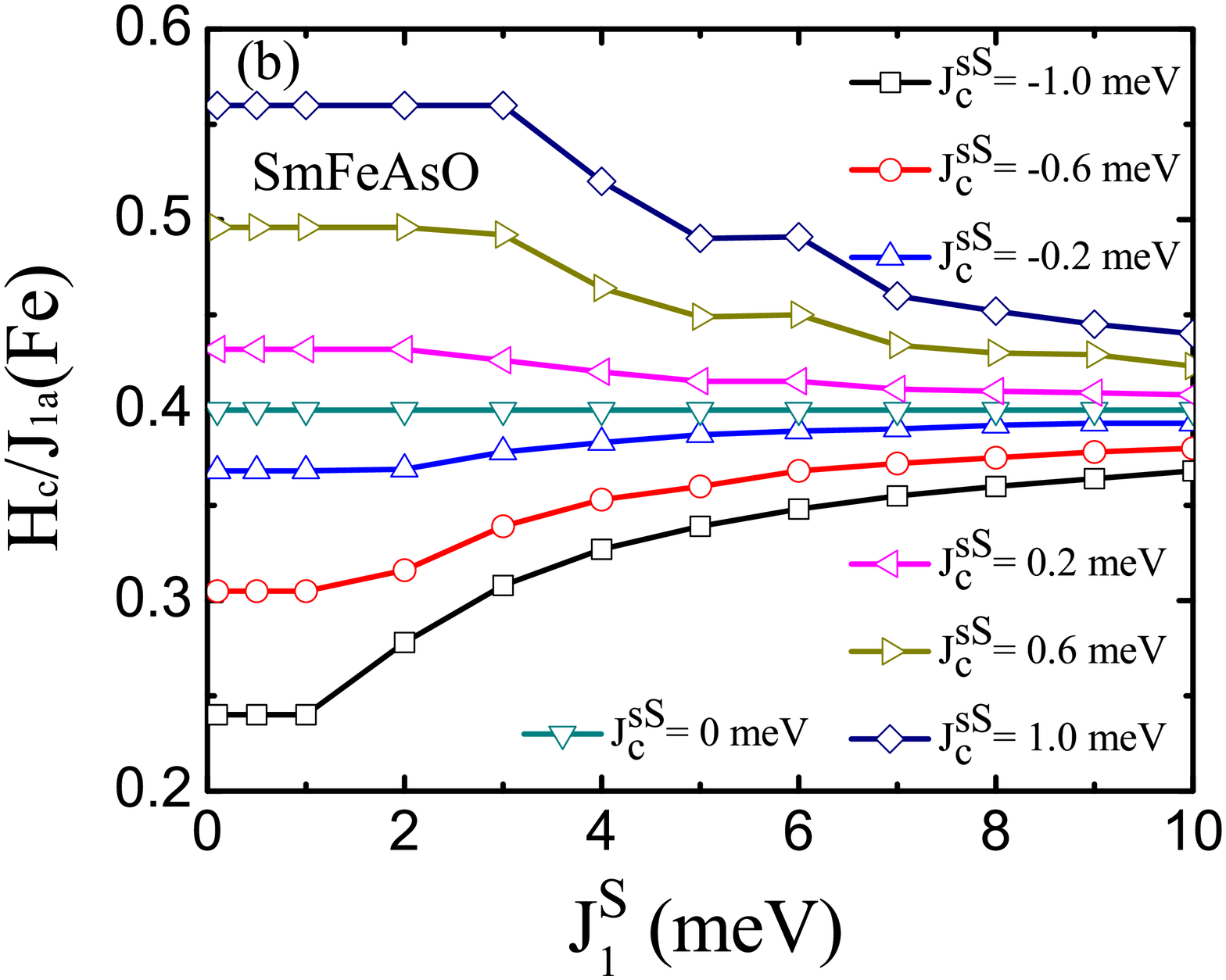}
\caption{Dependence of the critical magnetic field of the spin-flop transition for Fe$^{2+}$ ions on the interlayer coupling $J_{c}^{S}$ (a) and $J_{1}^{S}$ (b) in SmFeAsO under $H//a$.}
\label{Fig10}
\end{figure}

In SmFeAsO, the dependence of critical magnetic field of the Fe-spin flop ($H_{c}$(Fe)) on the interlayer coupling $J_{c}^{sS}$ and $J_{c}^{S}$ under $H//a$ is shown in Fig.~\ref{Fig10}(a), which exhibits similar tendency of EuFe$_{2}$As$_{2}$ case. The only difference is that no an effective magnetic field like FM of Eu ions is needed to overcome due to the AFM configuration in the rare-earth layers. In contrast to the case of EuFe$_{2}$As$_{2}$, the intralayer coupling of rare-earth ions $J_{1}^{S}$ also significantly affects the critical magnetic field value of the Fe spin-flop transition, since additional magnetic field is needed to overcome the AFM coupling between R spins.
In the case of CeFeAsO, a large positive value, $J_{c}^{sS}$$\thicksim$ 3.79 meV, is found in the $\mu SR$ experiment~\cite{PRB80-094524}, implying a very large coupling between Ce and Fe ions.

In the series of RFeAsO, the experiments found that the N$\acute{e}$el transition temperatures for R=Ce, Pr and Sm, Gd and Tb layers as well as Fe layers, are T$_{N}^{R}$=4.4, 11, 4.66, 4.11 and 2.54 K, and T$_{N}^{Fe}$=137, 123, 138, 128 and 122 K~\cite{PRB80-094524,PRB80-224511}, respectively. These indicate that the intralayer magnetic exchange couplings are only slightly different in these rare-earth compounds. Meanwhile, the neutron scattering experiments find that the interlayer magnetic coupling between R and Fe spins ({\it i.e.} $J_{c}^{sS}$) in CeFeAsO is stronger than those in SmFeAsO, showing that the weak interlayer coupling in SmFeAsO favors the spin fluctuations in FeAs layers, hence more high superconducting transition temperature. Also we expect that a small magnetic anisotropic energy of rare-earth spins, $J_{S}$, favors the spin fluctuations of Fe spins, thus enhances the pairing force and T$_{c}$.

\section{Conclusions}
In summary, we investigate the magnetic phase transition behavior in both 1111 and 122 systems with 4$f$-electrons. Our results demonstrate that the interplay of 3$d$ and 4$f$ spins plays a key role in the magnetic field dependence of these iron-based superconductors with magnetic rare-earth ions. The magnetic rare-earth layers, like magnetic intercalated layers, can tune the spin flop transition of the square Fe lattice in iron-based compounds. We expect that further experiments of strong static or pulsed magnetic field could verify these field-induced magnetic phase transitions.

\section*{Acknowledgment}

This work was supported by the National Sciences Foundation of China under Grant Nos. 11104274, 11274310, 11474287, and the Fundamental Research Funds for the Central University under Grant No. 27R1310020A. The calculations were performed in Center for Computational Science of CASHIPS, the ScGrid of Supercomputing Center and Computer Net work Information Center of Chinese Academy of Science.

\end{document}